# A Performance Comparison of Network Simulators for Wireless Networks

Atta ur Rehman Khan[a], Sardar M. Bilal[b], Mazliza Othman[a]

[a]Faculty of Computer Science & Information Technology, University of Malaya, LP 50603, Kuala Lumpur, Malaysia.
[b]Department of Telematic Engineering, University Carlos III de Madrid, 30 E-28911 Leganés, Madrid, Spain.

**Abstract:** *Network simulation is the most useful and common methodology used to evaluate different network topologies without real world implementation. Network simulators are widely used by the research community to evaluate new theories and hypotheses. There are a number of network simulators, for instance, ns-2, ns-3, OMNET++, SWAN, OPNET, Jist, and GloMoSiM etc. Therefore, the selection of a network simulator for evaluating research work is a crucial task for researchers. The main focus of this paper is to compare the state-of-the-art, open source network simulators based on the following parameters: CPU utilization, memory usage, computational time, and scalability by simulating a MANET routing protocol, to identify an optimal network simulator for the research community.*

## 1. Introduction

Wireless technology is advancing rapidly and new enhancements are proposed on a regular basis. In computer networks, new, untested protocols cannot be launched on a large scale due to uncertainty of its successful outcome. Therefore, the new protocols/schemes are tested with analytical modelling or simulation tools. After the simulation, if the new protocols show promising results, the protocols are implemented in the real world.

Analytical modelling has some drawbacks, for instance, the deduced results are not precise in terms of consumed energy, memory, and processing power. Although real world implementation provides realistic results, the physical implementation is a time consuming procedure and expensive as it may require a lot of hardware and human resources. Alternatively, simulation is affordable and provides good results in a cost effective way to evaluate the performance of proposed protocols/schemes.

In the early era, research work on communication networks involved both experimentation and mathematical modeling to prove feasibility and to establish bounds on expected performance of computer networks. However, in the last decade, computer networks have gone through a rapid revolution and have become too complicated for mathematical analysis. Computer-based simulation plays an important role in the research work to help the researchers and network designers to understand the behavior and performance of the networks and its protocols. Computer simulation is often used to test the planned capacity of networks and to meet customer requirements. In addition, simulation is also used to explore a wide range of potential protocol designs through rapid evaluation and iteration [1]. However, different simulators require variable time, memory and computation power for evaluating proposed protocols/techniques.

This paper presents performance comparison of four network simulators, i.e. ns-2, ns3, OMNET++, and GloMoSiM, because these are open source and well known network simulators in the research community. The main focus of this paper is to assist researchers in choosing the most suitable simulators for their work in terms of memory usage, computation time and CPU utilization.

The rest of the paper is organized as follows. Section 2 presents the related work while the simulation tools are explained in section 3. The selected routing protocol is described in section 4. Section 5 presents the simulation results and analysis, and the conclusion is provided in section 6.

## 2. Related Work

Selecting the appropriate network simulator is a crucial task for researchers. There are researchers who have been testing different routing protocols [2] in different simulators with different network parameters [3] to evaluate the precise performance of network protocols.
[5] compares ns-2 with OMNeT++ and QualNet by using radio propagation models. The architecture of ns-2 and TOSSIM are compared in [6]. In [7] and [8], the authors present performance comparison of different network simulators such as JavaSim, ns-2, and SSFNET. [9] specially focuses on simulators that are designed for sensor net-

works. [10] presents a qualitative comparison of ns-2 and OPNET. Furthermore, [11] presents a performance comparison of network simulators that are specially designed for VANETs.

The main difference between this paper and previously published papers is that we compare the most popular and open source simulators. Moreover, we perform the comparison by selecting the latest versions of simulators such as OMNET++ v4.2 and ns-3 v3.10. In addition, we use a MANET routing protocol, i.e. AODV, to evaluate the performance of the network simulators.

## 3. Simulation Tools

This section introduces the selected network simulators, i.e. ns-2, ns-3, OMNET++ and GloMoSiM. These simulators are selected because of their high popularity in the research community and most researchers use them to validate their new theories.

### 3.1) ns-2

Network Simulator-2 (ns-2) [23] is an open source, discrete event network simulator. It is used for the simulation of network protocols with different network topologies. It is capable of simulating wired as well as wireless networks. NS-2 was built in C++ and provides the simulation interface through OTcl, an object-oriented dialect of Tcl. The user describes a network topology by writing OTcl scripts, and then the main NS program simulates that topology with specified parameters. In ns-2, network animator (NAM) is used for the graphical view of the network. ns-2 is the most common and widely used network simulator for research work. NAM interface contains control features that allow users to forward, pause, stop and play the simulation. The interface of ns-2 is shown in Figure 1.

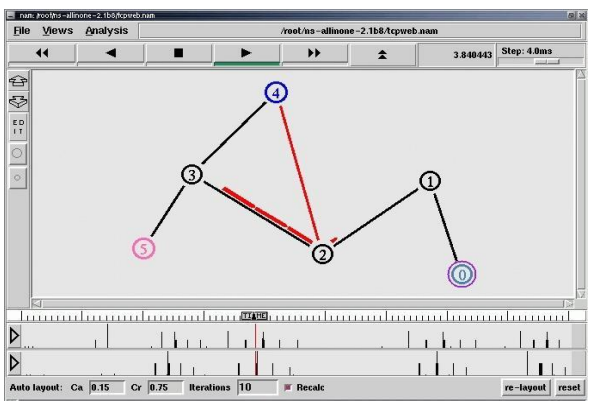

**Figure 1:** ns-2 Simulator Interface

In ns-2, arbitrary network topologies can be defined that are composed of routers, links and shared media [13]. The physical activities of the network are processed and queued in form of events, in a scheduled order. These events are then processed as per scheduled time that increases along with the processing of events. However, the simulation is not real time; it is considered virtual [12].

### 3.2) ns-3

The ns-3 project [24] was initiated in mid 2006 and is still under heavy development. Like ns-2, ns-3 is an open source, discrete-event network simulator. ns-3 is considered as a replacement of ns-2, not an extension [14]. Like ns-2, it does not have an OTcl API. It is written in C++ language and python. The latest version of ns-3 is ns-3.10 that supports parallel simulation and has an enhanced feature set. In addition, ns-3 network simulations can be implemented in pure C++, while some parts of the simulation can also be written using Python [7]. ns-3 interface is shown in Figure 2.

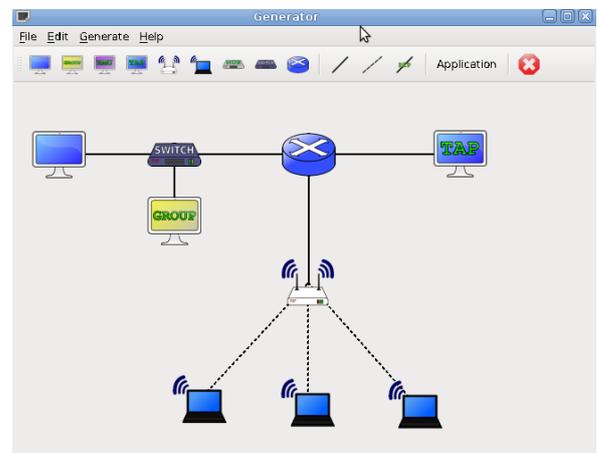

**Figure 2:** ns-3 Simulator Interface

ns-3 supports both simulation and emulation using sockets. It also generates pcap traces that can help in debugging. To analyze network traffic, standard tools like Wireshark [15] can be used to read trace files. ns-3 provides a realistic environment and its source code is well organized [16].

### 3.3) OMNET++

OMNET++ [25] has been available to the public since September 1997 and currently has a large number of users. Unlike ns-2 and ns-3, OMNET++ is not only designed for network simulations. It can be used for modeling of multipro-cessors, distributed hardware systems and performance evaluation of complex software systems. However, it is most commonly used for computer networks simulation. OMNET++ is a general discrete event, component-based (modular) open architecture simulation framework. OMNET++ interface is shown in Figure 3.

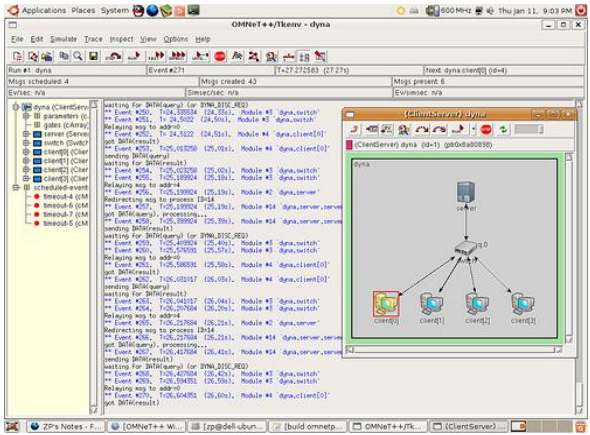

**Figure 3:** OMNET++ Simulator Interface

The motivation behind the development of OMNET++ was to produce a powerful open-source discrete event simulation tool that can be used by academic, educational and research-oriented commercial institutions for the simulation of computer networks, distributed and parallel systems [17]. OMNET++ distributions are available for both UNIX and Windows-based systems. It was developed using component-oriented approach that promotes structured and reusable models. In addition, OMNET++ has extensive graphical user interface (GUI) and intelligence support [18].

### 3.3) GloMoSiM

Global Mobile Information System Simulator (GloMoSiM) is a simulation environment used for large scale wireless networks. GloMoSiM uses parallel discrete-event simulation based on Parsec [19]. In addition, GloMoSiM uses the Parsec compiler to compile the simulation of protocols [20].

GloMoSiM is capable of simulating a network that contains thousands of nodes and heterogeneous communication links, for instance multicast and asymmetric links. In addition, GloMoSIm supports direct satellite communication, multi-hop wireless communication, and most of the traditional Internet protocols. GloMoSiM is a library-based sequential and parallel simulator that is designed purely for wireless networks [20].

GloMoSiM has a scalable simulation library that is based on the Parsec simulation environment [22].

It is developed as a set of library modules, each of which simulates a specific wireless communication protocol in the protocol stack [21].

## 4. Selecting a Routing Protocol

The ad hoc on demand distance vector (AODV) routing algorithm is used to compare the performance of the simulators. AODV is selected because of its pre-availability in the selected network simulators. AODV is a reactive routing protocol that establishes the route on demand (when the route is required by the source node) and maintains the route as long as required by the source node. It avoids the count to infinity problem by using sequential numbers on route updates. In addition, AODV maintains time-based states at each node and discards (expires) routing entries that are not recently used. AODV is relatively fast in terms of topological network changes and updates only the nodes that are affected by topological changes.

### 4.1. AODV

In AODV, the network remains silent unless a connection is needed by any node in the network. When a connection is required, the source node broadcasts a connection request. Referring to Figure 4, node A wants to communicate with node H, therefore, node A broadcasts a route request (RREQ) message in the network.

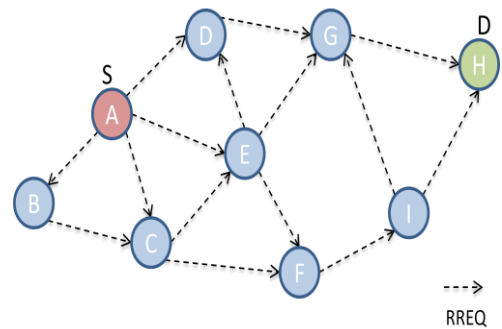

**Figure 4:** RREQ in AODV

Every node in the network forwards the RREQ message to its neighbors and records the previous node from where the request was received. When the destination node H receives the RREQ message, it sends back a unicast route reply (RREP) message to the source node through the node that delivered the RREQ message. The intermediate nodes that receives the RREP message forwards it to the next node with the smallest distance towards the source node as shown in Figure 5. The entries that are not used in the routing tables are recycled after a time. If a link fails, a routing error message is sent back to the transmitting node and the route discovery process is repeated.

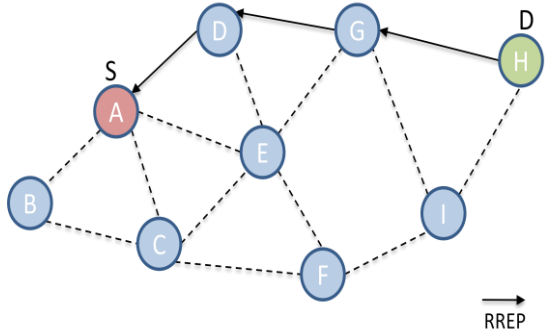

**Figure 5:** Route Maintenance Phase

The advantage of AODV is that it generates no extra traffic for communication along existing links. In addition, distance vector routing is simple and does not require much memory or calculation. However, AODV requires more time to establish a connection, and the initial communication to establish a route is higher than some other protocols.

The main advantage of this protocol is that the routes are established on demand, and destination sequence numbers are used to find the latest route towards the destination. The disadvantage of this protocol is that intermediate nodes can lead to inconsistent routes if the source sequence number is very old. If the intermediate nodes do not have the latest destination sequence number, stale entries may exist. In addition, multiple RouteReply packets may be created in response to a single RouteRequest packet that can lead to high control packet overhead. Moreover, AODV periodic beaconing leads to unnecessary bandwidth consumption.

## 5. Simulation and Results

To evaluate the performance of state of the art simulators, we simulated AODV routing protocol on selected simulators and evaluated the performance.

### 5.1. Simulation Setup

Before the start of a simulation, we configure connection establishments between pre-determined nodes, e.g. node A sends data to node B. As the communication starts, the source node starts transmitting at a regular interval of 0.2 seconds. During the simulation, the number of nodes was varied from 400 to 2000. In addition, each simulation was executed for 500 seconds in a simulation area of 1000 * 1000 (X * Y). The parameters are summarized in Table 1.

**Table 1:** Simulation setup

| SIMULATION/SCENERIO | |
|---|---|
| Simulation Time | 500s |
| X, Y Dimensions | 1000 x 1000 |
| Mobility Model | None |
| Packet size | 512 kb |
| Number of nodes | 400-2000 |
| Routing protocol | AODV |
| SIMULATORS VERSIONS | |
| ns-2 version | 2.34 |
| ns-3 version | 3.10 |
| GloMoSim version | 2.03 |
| OMNET++ version | 4.2 |

The simulations tools were executed on the Linux platform.

### 5.2. Results and Discussion

The performance comparison was done based on the following parameters: memory usage (MB), CPU utilization (percent), scalability, and computation time (seconds).

#### 5.2.1. Memory Usage

We simulated the AODV protocol for 500 seconds while varying the number of nodes from 400 to 2000. As shown in the Figure 6, ns-2 uses the highest amount of memory while ns-3 uses the lowest amount of memory compared to OMNET++ and GloMoSim. As the number of nodes increases, there is a linear growth in memory consumption for all simulators with minor difference. ns-3 was found to be the most efficient in memory usage among selected simulators.

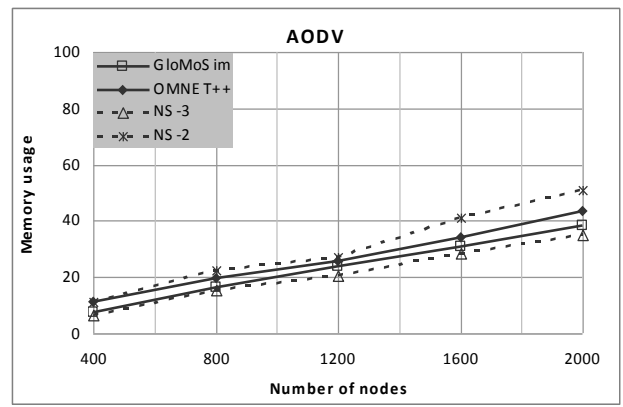

**Figure 6:** Number of nodes vs Memory usage

#### 5.2.2. CPU Utilization

CPU utilization was measured while varying the number of nodes. For all simulators, there is little effect on CPU utilization as the number of nodes increases.

Figure 7 shows that the CPU utilization of ns-2 and ns-3 is almost similar (5% variation) and is much higher compared to GloMoSiM and OMNET++.

GloMoSiM and OMNET++ shows a CPU utilization of only up to 35% with very little difference between them. The behavior of ns-2 and ns-3 was analyzed based on CPU utilization in detail by executing different applications in parallel with simulation tool. The simulations usually take a long time to execute while researchers use other applications, waiting for the results. We found that when other applications are executed in parallel, the CPU utilization of ns-2 and ns-3 drops to about 50%, hence allowing other applications to execute in parallel.

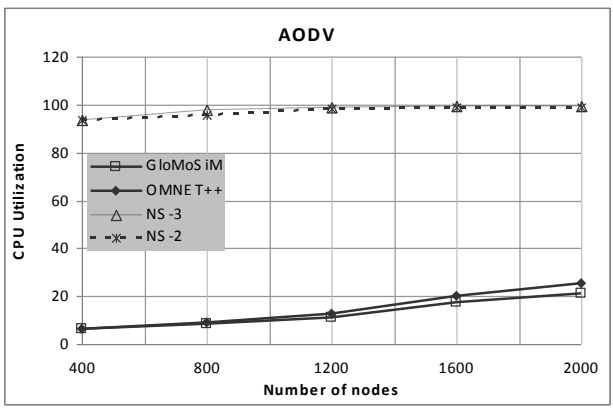

**Figure 7:** Number of nodes vs CPU Utilization

#### 5.2.3. Computation Time

The computation time was calculated by simulating AODV protocol for 500 seconds while increasing the number of nodes.

As illustrated in Figure 8, ns-2 has the highest computation time. In addition, ns-2's computation time increases rapidly with increasing number of nodes, which means ns-2 is not scalable. For large number of nodes, it may take a very long time compared to the other simulators. The computation time of the other simulators is quite low compared to ns-2. In terms of computation time and scalability, ns-3 appears to be the most efficient.

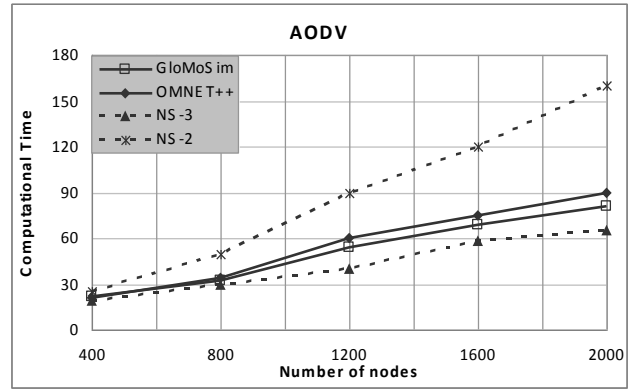

**Figure 8:** Number of nodes vs Computational Time

## 6. Conclusion

In this paper, we evaluate the performance of four network simulators with respect to different parameters. Based on the simulation results, we conclude that ns-3, OMNET++, and GloMoSiM are capable of carrying out large scale network simulations. ns-3 has proven to be the fastest simulator among the selected simulators in terms of computation time. In addition, ns-2 and ns-2 fully utilize the CPU, but is able to reduce CPU utilization when other applications are executed in parallel. Despite being quite new and is still under development, ns-3 demonstrates the best performance among all.

### References

[1] J. Heidemann, K. Mills, S. Kumar, "Expanding Confidence in Network Simulation," *IEEE Network Magazine,* Vol. 15, No. 5, Apr. 2000, pp. 58-63.

[2] S. R. Das, C.E. Perkins, E.M. Royer, "Performance comparison of two on-demand routing protocols for ad hoc networks," *INFOCOM,* Tel-Aviv, Israel, Aug. 2000.

[3] A. R. Khan, S. A. Madani, K. Hayat, S. U. Khan, "Clustering-based power-controlled routing for mobile wireless sensor networks," International Journal of Communication Systems, Vol. 25, No. 4, pp. 529-542, 2012.

[4] J. Broch, D. A. Maltz, D. B. Johnson, Y.C. Hu, J. Jetcheva, "A Performance Comparison of Multi-Hop Wireless Ad Hoc Network Routing Protocols,"
*ACM/IEEE international conference on Mobile computing and networking*, 1998.

[5] K. M Reineck, K. Jonas, K. Uhde, "Evaluation and Comparison of Network Simulation Tools" *Master Thesis*, Aug. 2008.

[6] Karl M., "A Comparison of the architecture of network simulators NS-2 and TOSSIM" *In Proc.*

*of Performance Simulation of Algorithm and Protocols Seminar Institute of parallel and distributed systems*, January 2005

[7] E. Weingarten, H. V. Lehn, K. Wehrle, "A performance comparison of recent network simulators," *IEEE International Conference on Communications,* 2009.

[8] D. M. Nicol, "Comparison of Network Simulators Revisited," *Dartmouth College*, http://www.ssfnet.org/Exchange/gallery/dumbbell/dumbbellperformance-May02.pdf, May 2002, accessed Jan 25[th], 2012.

[9] H. Sundani, H. Li, V. K. Devabhaktuni, M. Alam, "Wireless Sensor Network Simulators A Survey and Comparisons," *International Journal of Computer Networks*, Vol. 2, Apr 2011, pp. 249-265.

[10] B. Schilling, J. Hahner, "Qualitative Comparison of Network Simulation Tools, Modeling and Simulation of Computer Systems" *Technical report, Institute of Parallel and Distributed Systems,* University of Stuttgart, January 2005.

[11] A. Hassan, "VANET Simulation," *Masters Thesis in Electrical Engineering, School of Information Science, Computer and Electrical Engineering, Halmstad University,* May 2009.

[12] P. Yinfei, "Design Routing Protocol Performance Comparision in NS2: AODV Comparing to DSR as Example", *Department of Computer Science*, SUNY Binghamton, Vestal, 2010.

[13] C. H. Vrije, M. Hofmann, "Performance Comparison of Reliable Multicast Protocols using the Network Simulator ns-2" *Annual conference on local computer networks*, 1998.

[14] G. Carneiro, P. Fortuna, M. Ricardo, "FlowMonitor - a network monitoring framework for the Network Simulator 3 (NS-3)," *Fourth International ICST Conference on Performance Evaluation Methodologies and Tools*, 2009.

[15] Wireshark. http://www.wireshark.org, accessed Jan 10[th] 2012.

[16] T. Saeed, H. Gill, Q. Fei, Z. Zhang, B.T. Loo, "An Open-source and Declarative Approach towards Teaching Large-scale Networked Systems Programming," *ACM SIGCOMM Education Workshop, Toronto, Canada*, 2011.

[17] A. Varga, "THE OMNET++ DISCRETE EVENT SIMULATION SYSTEM" *European Simulation Multiconference* , 2003.

[18] A. Varga, A. Y. Sekercioglu, "Parallel Simulation Made Easy With OMNeT++" *15th European Simulation Symposium,* 2003.

[19] R. Bagrodia, R. Meyerr, "PARSEC: A Parallel Simulation Environment for Complex System" *UCLA technical report*, 1997.

[20] GloMoSiM, http://pcl.cs.ucla.edu/projects/glomosim/, accessed Sept 20[th] 2011.

[21] X. Zeng, R. Bagrodia, M. Gerla, "GloMoSiM: A Library for parallel simulation of Large scale wireless networks" *Workshop on Principles of Advanced and Distributed Simulation, pp. 154-161,* 1998.

[22] L. Bajaj, M. Takai, R. Ahuja, K. Tang, R. Bagrodia, M. Gerla, "GloMoSiM: A Scalable Network Simulation Environment," *IEEE Transactions on Reliability - TR*, 1999

[23] The network simulator, http://www.isi.edu.nsnam/ns, accessed Sept 20[th] 2011.

[24] T. R. Henderson, S. Roy, S. Floyd, and G. F. Riley, "ns-3 project goals," *workshop on ns-2: the IP network simulator*, ACM, New York, USA, 2006.

[25] A. Varga, R. Hornig, "An overview of the OMNeT++ simulation environment," *International Conference on Simulation Tools and Techniques for Communications, Networks and Systems*, Mar. 2008.